\documentclass[10pt]{article}
\usepackage{epsfig}
\begin{document}
\title{\bf Asymptotic behaviour of multiple scattering on infinite number
  of parallel demi-planes}
\author{E. Bogomolny,  and C. Schmit\\
 Laboratoire de Physique Th\'eorique et Mod\`eles Statistiques\\
Universit\'e de Paris XI, B\^at. 100\\
91405 Orsay Cedex, France}

\maketitle
\begin{abstract}

The exact solution  for the scattering of electromagnetic waves on an 
infinite number of parallel demi-planes has been obtained by J.F. Carlson
and A.E. Heins in 1947 using the Wiener-Hopf method \cite{CH}. 
We analyze their solution in the semiclassical limit of small wavelength and 
find the asymptotic behaviour of the reflection and transmission coefficients.
The results are compared with the ones obtained within the Kirchhoff
approximation.
\end{abstract}

\pagebreak

\section{Introduction}

For most systems classical mechanics corresponds to the $\hbar \to 0$ limit
of quantum mechanics and the nature of classical motion to a large extent
determines  spectral statistics of quantum problems \cite{Berry},
\cite{Bohigas}. Nevertheless, for certain models this correspondence breaks
down because limiting classical Hamiltonian has singularities and there
is no unique way to continue classical trajectories which hit them.
Though classical mechanics itself is not complete, quantum 
mechanics in many cases smoothes singularities and associates with
each of them a diffraction coefficient which gives a probability amplitude
for the scattering to a given channel. The simple example of such models is
given by diffractive models with small-size impurities where the total wave 
function at  large distances from an impurity is a sum of the free wave 
function plus a reflected field. In two dimension in a convenient normalization
\begin{equation}
\Psi(\vec{r}\ )=e^{i\vec{k}\vec{r}}+
\frac{D(\theta_f,\theta_i)}{\sqrt{8\pi k r}}e^{ikr-3i\pi/4}.
\label{standard}
\end{equation}
$D(\theta_f,\theta_i)$ is the diffraction coefficient for the
scattering of the incident plane wave of the direction $\theta_i$ 
to a reflected plane wave of the direction $\theta_f$.

Though in classical mechanics such scatters are negligible, 
their presence perturbs greatly (and in a calculable way) quantum mechanical 
problems (see \cite{Giraud} and references therein). 

An important class of diffractive systems consists of plane polygonal
billirds (see e.g. \cite{Richens}) with billiard corners playing the role of
diffraction centers. The main difficulty in  such type of diffractive  models
is the impossibility to represent everywhere the scattering field as a sum
of a free motion plus small corrections as in Eq.~(\ref{standard}) 
which forms  the basis of the usual scattering theory. 

\begin{figure}
\begin{center}
\epsfig{file=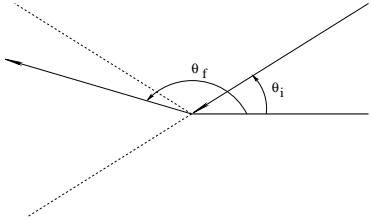, width=5cm, angle=-90 }
\end{center}
\caption{Scattering on the demi-plan (thick
  line). The initial and final rays are indicated by
  thin lines. The dotted lines represent optical boundaries.}
\label{plan}
\end{figure}

As an example we present the exact results for the diffraction on a demi-plan
derived by A. Sommerfeld in 1896  \cite{Sommerfeld2}. 
The total wave function for this problem obeying the  Helmholtz equation
\begin{equation}
(\Delta +k^2)\Psi (z,x)=0
\end{equation}
in the plan $(z,x)$ with  the Dirichlet boundary conditions along the
semi-infinite screen $x\geq 0$ has the following form (in notations of 
Fig.~\ref{plan})

\begin{eqnarray}
\Psi(\vec{r}\ )&=&e^{-ikr \cos (\theta_f-\theta_i)}
F(-\sqrt{2kr}\cos\frac{\theta_f-\theta_i}{2})\nonumber\\
&-&
e^{-ikr \cos (\theta_f+\theta_i)}
F(-\sqrt{2kr}\cos\frac{\theta_f+\theta_i}{2}),
\label{Sommerfeld}
\end{eqnarray}
where $F(u)$ is the Fresnel integral
\begin{equation}
F(u)=\frac{e^{-i\pi/4}}{\sqrt{\pi}}\int_{u}^{\infty}e^{it^2}dt.
\end{equation}
From the expansion of $\Psi(\vec{r}\ )$ at large distance it follows 
(see \cite{Sommerfeld2}) that the diffraction coefficient for this problem is
\begin{equation}
D(\theta_f,\theta_i)=  \frac{1}{\cos \frac{\theta_f-\theta_i}{2}}
-\frac{1}{\cos \frac{\theta_f+\theta_i}{2}}.
\label{diffraction}
\end{equation}
The important specificity of this diffraction coefficient is that it formally 
blows up in two directions 
\begin{equation}
\theta_f=\pi \pm \theta_i
\label{optbound}
\end{equation}
called optical boundaries (see Fig.~\ref{plan}) which separate regions 
with different number of 
geometrical optic rays which play the role of classical trajectories. As
wave fields are continuous, the separation of the exact field
(\ref{Sommerfeld}) in the sum of  free motion
plus small reflected fields is not possible in a vicinity of optical
boundaries which manifests itself as the divergence of the diffraction
coefficient (\ref{diffraction}). 

In practice one can use diffractive coefficient description for all scattering 
angles except intermediate parabolic regions near optical boundaries where the 
dimensionless arguments of $F$-functions in Eq.~(\ref{Sommerfeld}) are of 
the order of 1
\begin{equation}
u=\sqrt{kr}\sin \frac{\delta \varphi}{2}\sim 1,
\label{intermediate}
\end{equation}
and $\delta \varphi$ is the deviation angle from the optical boundaries
(\ref{optbound}). 

Difficult problems appear when inside these intermediate regions there are
new points of singular diffractions which is inevitable e.g. for plane polygonal
billiards.  In the semiclassical limit $k\to \infty$ certain cases of  
multiple diffraction on singular corners have been computed within the Kirchhoff
approximation in \cite{BPS}. The long-range nature of the 
diffraction on sharp corners leads to a strong interaction between different
singular points and no  general formulas for exact diffraction coefficients 
in these cases are known. 

In Refs.~\cite{CH} and \cite{HC} the exact solution for the reflection of 
electromagnetic wave on an infinite number of parallel semi-infinite metallic
sheets (equivalent to the scalar Helmholtz equation with the Dirichlet 
\cite{CH} and Neumann \cite{HC} boundary conditions) has been obtained by 
the Wiener-Hopf method (see also \cite{Noble}). Exact expressions for 
reflection and transmission coefficients (Eqs.~(\ref{R0}), 
(\ref{Rn}), and (\ref{Tm}) below) are quite cumbersome. They are given by infinite
products of different terms each depending non-trivially on the initial
momentum $k$. The purpose of this paper is the investigation of these results
in the semiclassical limit $k\to \infty$.

The plan of the paper is the following. In Section~\ref{general} general
properties of the reflection from periodic chain of demi-plans are
discussed. In Section~\ref{exactres} the exact expressions for diffraction
and transmission coefficients obtained in \cite{CH} are given and more
tractable expressions for the modulus of these coefficients are presented.
The most interesting case corresponds to the scattering with small incident
angle when in the intermediate region (\ref{intermediate}) there are many
singular points. In Section~\ref{elastic} two first terms of the expansion of
the elastic reflection coefficient are derived. In Section~\ref{smalln} the
first terms of expansion of the reflection coefficients for the forward scattering
in the power of the incident angle are calculated and in
Section~\ref{largen} the same is done for large-angle scattering. In
Section~\ref{finiteu} these calculations are generalized for all powers of
the incident angle provided the condition (\ref{intermediate}) is fulfilled.
The exceptional case when demi-plans are perpendicular to the scattering
plan is treated in Section~\ref{pi2}. In Section~\ref{kirchhoff} the exact
asymptotics are compared with the ones calculated within the Kirchhoff
approximation and in Section~\ref{summary} the summary of obtained results
is given. 

\section{Generalities}\label{general}

\begin{figure}
\begin{center}
\epsfig{file=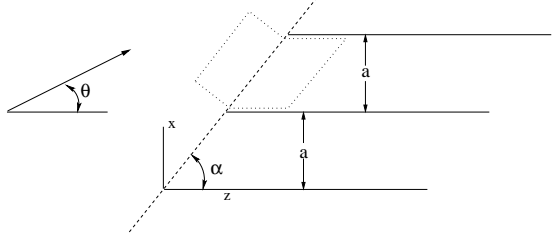, width=7cm, angle=-90 }
\end{center}
\caption{Scattering on  half-planes staggered by an angle $\alpha$. 
  $\theta$ is the angle of incidence of the initial plane wave. Dashed line
  indicates the scattering plane. Dotted line is the contour used in the
  computation of the current conservation.}
\label{fig3}
\end{figure}

The configuration of half-planes considered in Refs.~\cite{CH}, \cite{HC}
is represented  sche\-ma\-ti\-cally at Fig.~\ref{fig3}.
The equally spaced half-planes are parallel to the $z$-axis and their left
corners form a scattering plane inclined by an angle $\alpha$. 
The initial incident plane wave $\Psi_0(z,x)$ comes from the left with angle
of incidence $\theta$ with respect to $z$-axis. 
\begin{equation}
\Psi_0(z,x)=e^{ik(z\cos \theta+x\sin \theta)}.  
\end{equation}
In the far field due to the periodicity of the problem only finite number of
waves can propagate. The reflected plane waves have the following form
\begin{equation}
\Psi_n^{(ref)}(z,x)=e^{ik(z\cos \theta_n' +x\sin \theta_n')},
\label{reflectwave}
\end{equation}
where the allowed values of reflected angle 
$\theta_n'$ are determined by the usual grating equation (see e.g.
\cite{Sommerfeld2})
\begin{equation}
kd(\cos \varphi-\cos \varphi_n')=2\pi n.
\label{grating}
\end{equation}
Here $\varphi$ and $\varphi_n'$ are the incident  and  reflected angles 
defined as the angles between the scattering plane 
and  the incident  (respectively, reflected) plane wave direction 
\begin{equation}
\varphi=\alpha-\theta,
\end{equation}
and
\begin{equation}
\varphi_n'=\theta_n'-\alpha.
\end{equation}
$d=a/\sin \alpha$ is the distance along the scattering plane between corners of
the half-planes, $n$ is an integer.

In terms of the dimensionless momentum
\begin{equation}
Q=\frac{kd}{\pi}
\label{Q}
\end{equation}
Eq.~(\ref{grating}) determines a finite number of possible reflected
directions when $\varphi$ is fixed 
\begin{equation}
\cos \varphi_n'=\cos \varphi -\frac{2n}{Q}.
\end{equation}
The allowed values of $n$  are restricted by the inequalities followed from 
this expression
\begin{equation}
-Q\sin^2\frac{\varphi}{2}\leq n \leq Q\cos^2\frac{\varphi}{2}.
\end{equation}
For later use it is convenient to introduce instead of the initial angle 
$\varphi$   a  new variable 
\begin{equation}
u=\sqrt{Q}\sin \frac{\varphi}{2}
\label{u}
\end{equation}
which is suitable for the description of intermediate regions
(\ref{intermediate}) where the usual diffraction coefficient description can
not be applied. To investigate the
multiple scattering in such regions, we consider below the limiting case
$Q\to \infty$ with $u$ kept fixed which corresponds to small initial
angles of the order of $1/\sqrt{Q}$
\begin{equation}
\varphi\approx\frac{2u}{\sqrt{Q}}.
\label{phismall}
\end{equation}
For reflected waves we shall distinguish between small angle reflection where
the variable
\begin{equation}
u_n=\sqrt{Q}\sin \frac{\varphi_n'}{2}
\label{uprime}
\end{equation}
remains finite when $Q\to\infty$ and large angle reflection otherwise.

In terms  of  variables (\ref{u}) and (\ref{uprime}) the grating equation 
(\ref{grating}) reads
\begin{equation}
u_n=\sqrt{n+u^2}
\label{nu}
\end{equation}  
Physical values of reflection angles should obey the inequality
$0\leq \varphi_n'\leq \pi$ or $0\leq \sin \varphi_n'\leq 1$. From
(\ref{grating}) it follows that the physical branch is defined by the
condition 
\begin{equation}
\sin \varphi_n'=2\sqrt{\frac{n+u^2}{Q}
  \left (1-\frac{n+u^2}{Q}\right )}\geq 0.
\label{branch}
\end{equation}
In the notations of Fig.~\ref{fig3} the grating equation (\ref{grating})
takes the form used in \cite{CH}
\begin{equation}
k\rho -\omega_n b-a\sqrt{k^2-\omega_n^2}=2\pi n,
\label{omegan}
\end{equation}
where 
\begin{equation}
\rho=a\sin \theta +b\cos \theta=d\cos \varphi,
\end{equation}
$b=a/\tan \alpha$ is the shift of the half-plane corners along the $z$-axis,
and 
\begin{equation}
\omega_n=k\cos \theta_n'=k\cos (\alpha+\varphi_n')
\label{omegaref}
\end{equation}
is the projection  of   the final momentum on  the $z$-axis. 
The positive (resp.
negative) sign of $\sqrt{k^2-\omega_n^2}$ correspond to forward (resp.
backward) scattering.

In particular, the specular reflection corresponds to $n=0$ in the above
equation and 
\begin{equation}
\omega_0=k\cos (2\alpha-\theta)=k\cos (\alpha+\varphi).
\label{omega0}
\end{equation}

The possible transmission modes correspond to motion in a straight tube  
\begin{equation}
\Psi_m^{(trans)}(z,x)=e^{i\widetilde{\omega}_m z} \sin \frac{\pi}{a}mx ,
\label{transition}
\end{equation}
where the transmitted frequencies are
\begin{equation}
\widetilde{\omega}_m=\sqrt{k^2-\left (\frac{\pi m}{a}\right )^2}=
k\sqrt{1-\left (\frac{m}{Q'}\right )^2}.
\label{tildomega}
\end{equation}
Here
\begin{equation}
Q'=\frac{ka}{\pi}=Q\sin \alpha
\end{equation}
and $1\leq m\leq Q'$ is an integer.

\section{Exact results}\label{exactres}

At large distances the total reflected field is the sum over the individual
allowed propagating modes (\ref{reflectwave}) 
\begin{equation}
\Psi^{(ref)}(z,x)=\sum_{-Q\sin^2 (\varphi/2)\leq n\leq Q\cos^2 (\varphi/2)}
R_n e^{ik[z\cos (\alpha +\varphi_n')+x\sin (\alpha +\varphi_n')]},
\label{totalreflection}
\end{equation}
with $\varphi_n'$ determined by Eq.~(\ref{grating}) and the coefficients $R_n$
in the sum are reflection coefficients  which define the probability 
amplitudes of reflection to the given channel.

Similarly, the total transmitted field is the sum over individual
transmitted modes (\ref{transition}) 
\begin{equation}
\Psi^{(trans)}(z,x)=\sum_{1\leq m\leq Q\sin \alpha }
T_m e^{i\widetilde{\omega}_m z} \sin \frac{\pi}{a}mx 
\label{totaltransmision}
\end{equation}
with $\widetilde{\omega}_m$ from Eq.~(\ref{tildomega}) and the coefficients 
$T_m$ are called the transmission coefficients.

We restrict ourselves to the case of the Dirichlet boundary conditions on
the demi-plans treated in \cite{CH}. 
From the results of this work one can write down explicit expressions
for reflection and transmission coefficients in the form of  infinite products.

The reflection coefficients  in Eq.~(\ref{totalreflection}) with $n=0$ and $n\neq 0$
are slightly different. For $n=0$ the reflection coefficient $R_0$
corresponds to the specular reflection $\varphi_0=\varphi$ and we shall call
it the elastic reflection coefficient. From \cite{CH}  one gets
\begin{equation}
R_0=-\frac{K(\omega')}{K(\omega_0)}.
\label{R0}
\end{equation}
For non-zero $n$ \cite{CH}
\begin{equation}
R_n=\frac{(\omega' -\omega_0)}{(\omega_n-\omega')(\omega_n-\omega_0)}
\frac{K(\omega')}{K'(\omega_n)}.
\label{Rn}
\end{equation}
The transmission coefficients in the expansion (\ref{totaltransmision})  
are given by the following expressions \cite{CH}
\begin{equation}
T_m=-\frac{\pi m (\omega'-\omega_0)}{a^2 \widetilde{\omega}_m 
  (\widetilde{\omega}_m-\omega')(\widetilde{\omega}_m-\omega_0)}
\left [1-(-1)^me^{i(k\rho - \widetilde{\omega}_m b)}\right ]
  \frac{K(\omega')}{K(\widetilde{\omega}_m)}.
\label{Tm}
\end{equation}
In these formulas $\omega_0$ is given by (\ref{omega0}),
$\omega_n$ are defined by (\ref{omegan}) and 
\begin{equation}
\omega'=k\cos \theta=k\cos (\alpha -\varphi).
\end{equation}
The function $K(\omega)$ is the ratio of two functions
\begin{equation}
K(\omega)=\frac{g_{+}(\omega)}{f_{+}(\omega)}e^{\chi \omega},
\label{Komega}
\end{equation}
with
\begin{equation}
\chi=-i\frac{a}{\pi}[(\alpha-\frac{\pi}{2})\frac{1}{\tan \alpha} -\ln (2\sin\alpha)],
\end{equation}
The denominator  $f_{+}(\omega)$ in (\ref{Komega}) is given by  the convergent infinite product
\begin{equation}
f_{+}(\omega)=\prod_{n=1}^{\infty}
\left [\sqrt{1-\left (\frac{ka}{\pi n}\right )^2}-
  i\frac{\omega a}{\pi n}\right ]e^{i\omega a/\pi n}.
\label{fplus}
\end{equation}
The numerator $g_{+}(\omega)$ in (\ref{Komega}) is the product of two functions
\begin{equation}
g_{+}(\omega)=G_1(\omega)G_2(\omega),
\end{equation}
where $G_{1,2}(\omega)$ are represented as the following
convergent infinite products
\begin{equation}
G_1(\omega)=\prod_{n=1}^{\infty} (\Delta_n-i\Psi_n)
e^{(k\rho-\omega b+i\omega a)/(2\pi n)+i(\pi/2-\alpha) },
\end{equation}  
\begin{equation}
G_2(\omega)=\prod_{n=-\infty}^{-1} (\Delta_n+i\Psi_n)
e^{(k\rho-\omega b-i\omega a)/(2\pi n)-i(\pi/2-\alpha) },
\end{equation}
and
\begin{eqnarray}
\Delta_n&=&\sqrt{\sin^2\alpha \left (1-\frac{k\rho}{2\pi n}\right )^2-
  \left (\frac{ak}{2\pi n}\right )^2},\\  
\Psi_n&=&\frac{\omega a}{2\pi n \sin \alpha}+
\cos \alpha \left (1-\frac{k\rho}{2\pi n}\right ).
\end{eqnarray}
The functions $f_{+}(\omega)$ and $g_{+}(\omega)$  have no singularities or 
zeros in the upper half plane
\begin{equation}
\mbox{Im } \omega> \mbox{Im } \omega_0
\end{equation}
where one assumes that the momentum $k$ has a small positive imaginary part. 

These functions appear in the following Wiener-Hopf type factorization problem
\begin{equation}
f_{+}(\omega)f_{-}(\omega)=
\frac{\sin  [a\sqrt{k^2-\omega^2}]}{a\sqrt{k^2-\omega^2}},
\label{fpm}
\end{equation}
and
\begin{equation}
\frac{a^2+b^2}{2}(\omega-\omega_0)(\omega-\omega')g_{+}(\omega)g_{-}(\omega)=
\cos [a\sqrt{k^2-\omega^2}]-\cos [k\rho -\omega b].
\label{gpm}
\end{equation}
Functions $f_{-}(\omega)$ and $g_{-}(\omega)$ have no singularities in the
lower half plane 
\begin{equation}
\mbox{Im } \omega < \mbox{Im } \omega'.
\end{equation}
The explicit form of these functions is 
\begin{equation}
f_{-}(\omega)=\prod_{n=1}^{\infty}
\left [\sqrt{1-\left (\frac{k a}{\pi n}\right )^2}+i\frac{\omega a}{\pi
  n}\right ]e^{-ia\omega/\pi n},
\label{fminus}
\end{equation}
and
\begin{equation}
g_{-}(\omega)=\widetilde{G}_1(\omega)\widetilde{G}_2(\omega),
\end{equation}
where
\begin{equation}
\widetilde{G}_1(\omega)=\prod_{n=1}^{\infty} (\Delta_n+i\Psi_n)
e^{(k\rho-\omega b-i\omega a)/(2\pi n)-i(\pi/2-\alpha) },
\end{equation}  
\begin{equation}
\widetilde{G}_2(\omega)=\prod_{n=-\infty}^{-1} (\Delta_n-i\Psi_n)
e^{(k\rho-\omega b+i\omega a)/(2\pi n)+i(\pi/2-\alpha) }.
\end{equation}
The real zeros of the function $g_{+}(\omega)$ 
\begin{equation}
g_{+}(\omega_n)=0
\end{equation}
correspond to pure imaginary values of $\Delta_n$ and coincide with 
the reflected frequencies (\ref{omegaref}). Their explicit form is
\begin{equation}
\omega_n=k\cos \alpha \left (\cos \varphi -\frac{2\pi n}{kd}\right )-  
k\sin \alpha
\sqrt{\sin^2 \varphi +\frac{4\pi n}{kd}\left (\cos \varphi -\frac{\pi
    n}{kd}\right )}.
\label{omegakn}
\end{equation}
In accordance with Eq.~(\ref{branch}) the positive branch of the square root
has to be chosen.  
$\omega_n$ with positive  (resp. negative) $n$ are real zeros of
$G_1(\omega)$ (resp. of $G_2(\omega)$). 

For the later use we need also the zeros
$\omega_n^*$ of  the function $g_{-}(\omega)$. They are given by
Eq.~(\ref{omegakn}) with the different sign of the square root
\begin{equation}
\omega_n^*=k\cos \alpha \left (\cos \varphi -\frac{2\pi n}{kd}\right )+  
k\sin \alpha
\sqrt{\sin^2 \varphi +\frac{4\pi n}{kd}\left (\cos \varphi -\frac{\pi
    n}{kd}\right )}.
\label{omegaknn}
\end{equation}
The transmitted frequencies $\widetilde{\omega}_m$ are zeros of the function
$f_{-}(\omega)$. They are given by (\ref{tildomega}). The zeros of
$f_{+}(\omega)$ are just $-\widetilde{\omega}_m$.  

The expressions (\ref{R0}), (\ref{Rn}), and (\ref{Tm}) for the reflection
and transmission coefficients include infinite products and are quite
cumbersome. Simpler formulas can be obtained for the modulus of these
coefficients which is sufficient for many purposes. 
Using repeatedly the defining relations (\ref{fpm}) and (\ref{gpm}) one
can demonstrate that the values of $|R_n|^2$ and $|T_n|^2$ are expressed 
by the following finite products
\begin{eqnarray}
|R_0|^2&=&\left | \prod_{n'\neq 0} \frac{\omega_0-\omega_{n'}^*}{\omega_0-\omega_{n'}}
\right . 
\prod_{m'\neq 0}\left . 
\frac{\omega_0+\widetilde{\omega}_{m'}}{\omega_0-\widetilde{\omega}_{m'}}
\right | |K(\omega')|^2,
\label{modr0}\\
|R_n|^2&=&\left |\frac{4\sin^2\alpha \sin^2\varphi}
{(\omega_n-\omega')(\omega_n-\omega_0)}\right .\nonumber\\
&\times&
\prod_{\begin{array}{c}{\scriptstyle n'\neq 0}\\{\scriptstyle n'\neq
    n}\end{array}} 
\frac{\omega_n-\omega_{n'}^*}{\omega_n-\omega_{n'}}
\prod_{m'\neq 0}\left .
\frac{\omega_n+\widetilde{\omega}_{m'}}{\omega_n-\widetilde{\omega}_{m'}}
\right | |K(\omega')|^2,
\label{modrn}\\
|T_m|^2&=&\left |\frac{4 \sin^2\varphi}
{(\widetilde{\omega}_m-\omega')(\widetilde{\omega}_m-\omega_0)}\right .
\nonumber\\
&\times& 
\prod_{n'\neq 0} \frac{\widetilde{\omega}_m-\omega_{n'}^*}{\widetilde{\omega}_m-\omega_{n'}}
\prod_{\begin{array}{c}{\scriptstyle m'\neq 0}\\{\scriptstyle m'\neq m}
\end{array}}\left .
\frac{\widetilde{\omega}_m+\widetilde{\omega}_{m'}}
{\widetilde{\omega}_m-\widetilde{\omega}_{m'}}\right | |K(\omega')|^2,
\label{modtm}
\end{eqnarray}
and 
\begin{equation}
|K(\omega')|^2=\left | \prod_{n'\neq 0}
\frac{\omega'-\omega_{n'}}{\omega'-\omega_{n'}^*}
\prod_{m'\neq 0}
\frac{\omega'-\widetilde{\omega}_{m'}}{\omega'+\widetilde{\omega}_{m'}}\right |.
\label{modk}
\end{equation}
In Eds.~(\ref{modr0})-(\ref{modk})  the products are taken
over finite number of  real reflected (\ref{omegakn}) and transmitted
(\ref{tildomega}) frequencies.

As $\omega'-\omega_0=2\sin \alpha \sin \varphi$ and $\omega_0^*=\omega'$,
Eqs.~(\ref{modr0}) and (\ref{modrn}) can be rewritten in a form
which is valid for both $n=0$ and $n\neq 0$
\begin{equation}
|R_n|^2=\left |\left (\frac{\omega_0-\omega'}
{\omega_n-\omega'}\right )^2
\prod_{ n'\neq n} 
\frac{\omega_n-\omega_{n'}^*}{\omega_n-\omega_{n'}}
\prod_{m'\neq 0}\frac{\omega_n+\widetilde{\omega}_{m'}}{\omega_n-\widetilde{\omega}_{m'}}
\right | |K(\omega')|^2.
\label{commodrn}
\end{equation}
If $n\neq 0$ the first product includes the term with $n'=0$.

The scattering amplitudes fulfill the current conservation. In the
configuration of Fig.~\ref{fig3} it is convenient to consider the
current through the surface of a rectangle enclosing the scattering plane whose
two sides are parallel to it. Due to the periodicity of the demi-planes this
surface is reduced to the one indicated by dotted line in Fig.~\ref{fig3}
and the current through boundaries perpendicular to the scattering plane can
be ignored. In this case the total current through the parallel parts of 
this surface is zero
and a simple calculation gives the following relation between the reflected
and transmission coefficients
\begin{equation}
k\sin \varphi =k \sum_{n} \sin \varphi_n' \ |R_n|^2+
\sum_{m}\sin \alpha \ |T_m|^2\sqrt{k^2-\left (\frac{\pi m}{a}\right )^2}.
\label{current}
\end{equation}
All expressions in this Section are exact and suitable for numerical
calculations. But for theoretical purposes they are practically
intractable, especially in the semiclassical limit $Q\to \infty$, because the
number of factors in Eqs.~(\ref{modr0})-(\ref{modtm}) increases with $Q$ and
each factor depends non-trivially on $Q$. 

In the next Sections we investigate these expressions  when  $Q\to \infty$.  
In this limit non-trivial results appear  when the incident 
wave forms a small angle  $\varphi$ with the scattering plane (as in
(\ref{phismall})) and we focus our attention on this region.

\section{Elastic reflection coefficient}\label{elastic}

Let us consider first the behavior of the elastic reflection coefficient 
$R_0$ in Eq.~(\ref{R0}) in the limit $\varphi \rightarrow 0$. Taking into account
only the linear in  $\varphi$ terms direct calculations give
\begin{equation}
R_0=-(1+ \beta \varphi)
\end{equation}
and
\begin{eqnarray}
&&\beta  = i\sum_{n=1}^{\infty} \xi_n \left [
\frac{2 \sin  \alpha}{\sqrt{(1-\xi_n^2)}-i\xi_n\cos \alpha }-
\frac{1}{\sqrt{(\sin^2\alpha +\xi_n\sin \alpha )} +i \cos \alpha }\right .
\nonumber\\
&&-\left .\frac{1}{\sqrt{(\sin^2\alpha -\xi_n \sin \alpha )}-i \cos \alpha}
\right ]
\nonumber\\
&&-2i Q'\sin \alpha  
[(\alpha-\frac{\pi}{2})\frac{1}{\tan \alpha} -\ln (2\sin\alpha)],
\label{61}
\end{eqnarray}
where $\xi_n=Q'/n$ and $Q'=ka/\pi=Q\sin \alpha$.

Separating  the real and imaginary parts of Eq.~(\ref{61}) one obtains
\begin{equation}
\beta=Q'(C_1+iC_2),
\label{betaprime}
\end{equation}
where
\begin{eqnarray}
C_1&=&-2\sin \alpha \sum_{n=1}^{[Q']}\frac{1}{\sqrt{Q'^{2}-n^2}+Q'\cos \alpha}+
\sum_{n=1}^{[Q]}\frac{1}{\sin \alpha \sqrt{(Q-n)n}+n\cos \alpha}
\nonumber \\
&-&\cos \alpha \sum_{n=1}^{[Q']}\frac{1}{n+Q'\sin \alpha }
-\cos \alpha \sum_{n=[Q']+1}^{[Q]} \frac{1}{n-Q'\sin \alpha},
\label{C1}
\end{eqnarray}
and
\begin{eqnarray}
C_2&=&\sin \alpha \left [ \sum_{n=[Q']+1}^{\infty}
\frac{\sqrt{n^2-Q'^2}}{n^2-Q'^2\sin^2\alpha}-
\sum_{n=1}^{\infty}\frac{\sqrt{n+Q}}{\sqrt{n}(n+Q'\sin \alpha)}\right .
\nonumber \\
&-&\left .\sum_{n=[Q]+1}^{\infty}\frac{\sqrt{n-Q}}{\sqrt{n}(n-Q'\sin \alpha)} 
-2[(\alpha-\frac{\pi}{2})\frac{1}{\tan \alpha} -\ln (2\sin\alpha)]\right ]. 
\label{C2}
\end{eqnarray}
Here and below $[x]$ denotes the integer part of $x$ such that $x=[x]+\{x\}$
and the fractional part $\{x\}$ obeys the inequality $0\leq \{x\}<1$.

In the computations below we assume that $\alpha \neq 0$, $\pi/2$, and $\pi$.
In the semiclassical limit $k\rightarrow \infty$ all sums
in the above expressions include many terms which  can be estimated by
usual methods. The dominant contribution corresponds to the change of the
summation to the integration. One gets
\begin{eqnarray}
\widetilde{C}_1&=&-2\sin \alpha \int_0^{Q'}\frac{dn}{\sqrt{Q'^2-n^2}+Q'\cos \alpha}+
\int_{0}^{Q}\frac{dn}{\sin \alpha \sqrt{(Q-n)n}+n\cos \alpha}
\nonumber \\
&-&\cos \alpha \int_{0}^{Q'}\frac{dn}{n+Q'\sin \alpha }
-\cos \alpha \int_{Q'}^{Q} \frac{dn}{n-Q'\sin \alpha}
\end{eqnarray}
and
\begin{eqnarray}
\widetilde{C}_2&=&\sin \alpha \left [ \int_{Q'}^{\infty}
\frac{\sqrt{n^2-Q'^2}}{n^2-Q'^2\sin^2\alpha}dn-
\int_{0}^{\infty}\frac{\sqrt{n+Q}}{\sqrt{n}(n+Q'\sin \alpha)}dn\right .
\nonumber \\
&-&\left .\int_{Q}^{\infty}\frac{\sqrt{n-Q}}{\sqrt{n}(n-Q'\sin \alpha)}dn 
-2[(\alpha-\frac{\pi}{2})\frac{1}{\tan \alpha} -\ln (2\sin\alpha)]\right ]. 
\end{eqnarray}
The integrals are elementary and cancel each other, i.e.
$\widetilde{C}_1=\widetilde{C}_2=0$. One can prove that
the first correction term (of the order of
$1/\sqrt{Q}$) to $C_1$ appears from small $n$ summation in the second term in 
Eq.~(\ref{C1})
\begin{eqnarray}
C_1&\rightarrow&
\sum_{n=1}^{[Q]}\frac{1}{\sin \alpha \sqrt{(Q-n)n}+n\cos \alpha}
+\mbox{smooth terms}\nonumber\\
&\rightarrow&
\frac{1}{\sin \alpha\sqrt{Q}}\lim_{N\to\infty}
\left (\sum_{n=1}^N \frac{1}{\sqrt{n}}-\int_0^N\frac{dn}{\sqrt{n}}\right )
\end{eqnarray}
The integral is subtracted because we know that all integrals over $n$ are
canceled by other terms.

The last limit can be computed from the relation
\begin{equation}
\lim_{N\to\infty}
\left (\sum_{n=1}^N \frac{1}{\sqrt{n}}-\frac{\sqrt{N}}{2}\right
  )=\zeta (\frac{1}{2})
\label{zeta}  
\end{equation}
where $\zeta(s)$ is the Riemann zeta function 
$(\zeta (\frac{1}{2})\approx-1.460354)$.

Finally when $Q\to \infty$
\begin{equation}
C_1\to \frac{1}{\sin \alpha \sqrt{Q}}\zeta (\frac{1}{2}).
\end{equation}
Similarly the dominant contribution to $C_2$  comes from small-$n$ summation
in  the second term of Eq.~(\ref{C2})
\begin{equation}
C_2\rightarrow   
-\sin \alpha \sum_{n=1}^{\infty}\frac{\sqrt{n+Q}}{\sqrt{n}(n+Q'\sin \alpha)}
+\mbox{smooth terms}\rightarrow 
-\frac{1}{\sin \alpha \sqrt{Q}}\zeta (\frac{1}{2}).
\end{equation}
From (\ref{betaprime}) one concludes that for large $k$ and 
$\alpha \neq 0, \pi/2, \pi$
\begin{equation}
\beta=\sqrt{Q}(1-i)\zeta (\frac{1}{2}).
\end{equation}
In terms of variable (\ref{u}) this result states that the two first terms
of the expansion of the elastic reflection coefficient into the power of $u$
are the following
\begin{equation}
R_0=-1-2\sqrt{2}e^{-i\pi/4}\zeta(\frac{1}{2})u.
\label{2uR0}
\end{equation}
In Section~\ref{kirchhoff} we demonstrate that  these two terms can be
obtained from the Kirchhoff approximation developed in \cite{BPS}.  

\section{Small angle reflection in the limit $\varphi\to 0$}\label{smalln}

When the incident angle $\varphi \rightarrow 0$ it follows from Eqs.~({\ref{Rn})
and (\ref{Tm})  that the reflection coefficients with $n\neq 0$ and all
transmission coefficients are proportional to $\varphi$.

In the computation below it is convenient to rescale all frequencies by $k$ 
i.e. to change
\begin{equation}  
\omega\to \frac{\omega}{k}.
\end{equation}
To simplify the notation  we  shall from now use  the same symbols for the 
rescaled quantities.

When $\varphi\to 0$ rescaled Eqs.~(\ref{omegakn}), (\ref{omegaknn}), and 
(\ref{tildomega}) give
\begin{equation}
\omega_n=\cos \alpha \left (1 -\frac{2 n}{Q}\right )-2\sin \alpha
\sqrt{\frac{n}{Q}\left (1 -\frac{n}{Q}\right )},
\label{omegasmall1}
\end{equation}
\begin{equation}
\omega_n^*=\cos \alpha \left (1 -\frac{2 n}{Q}\right )+
2\sin \alpha
\sqrt{\frac{n}{Q}\left (1 -\frac{n}{Q}\right )},
\label{omegasmall2}
\end{equation}
and 
\begin{equation}
\widetilde{\omega}_m=\sqrt{1-\frac{m^2}{ Q^2 \sin^2 \alpha}}
\label{tildesmall}  
\end{equation}  
with all square roots chosen positive.

Using  Eqs.~(\ref{modrn}) and (\ref{modk}) one  can compute the reflection 
coefficient  modulus in limit $k\rightarrow \infty$ by writing  the logarithm 
of each product as a sum over the corresponding frequencies. E.g.
$$
\log \prod_{n'\neq  n}|\omega_n-\omega_{n'}|= \sum_{n'\neq n}\log |\omega_n-\omega_{n'}|.
$$
When $k\rightarrow \infty$ the sum can be substituted by the integral.
Exactly as it was done in the precedent Section one can check that all
integrals in the full product (\ref{modrn}) cancel and the dominant 
contributions come from regions with small factors. 

We are interested first in the behaviour of the reflection
coefficients $R_n$ at fixed $n\neq 0$. In this case there are 3 regions with 
small factors. The first corresponds to 
$\omega_n-\widetilde{\omega}_{m'}\rightarrow 0$, the
second appears when $\omega_n-\omega_{n'}^*\rightarrow 0$, and the third
includes  cases when $n'$ is close to $n$ (see Fig.~(\ref{omega12})).
\begin{figure}
\begin{center}
\epsfig{file=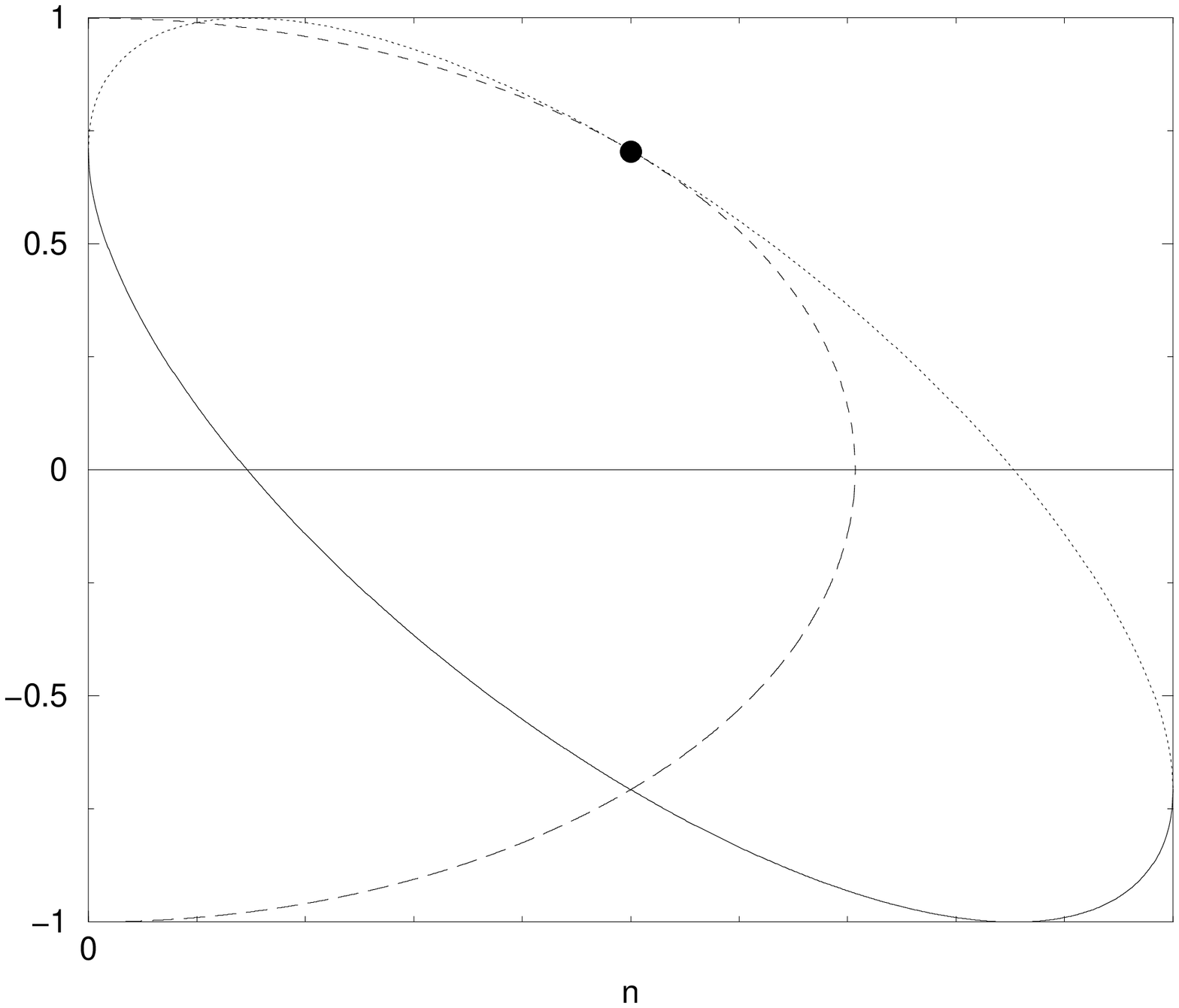, width=5cm, angle=-90 }
\hfill
\epsfig{file=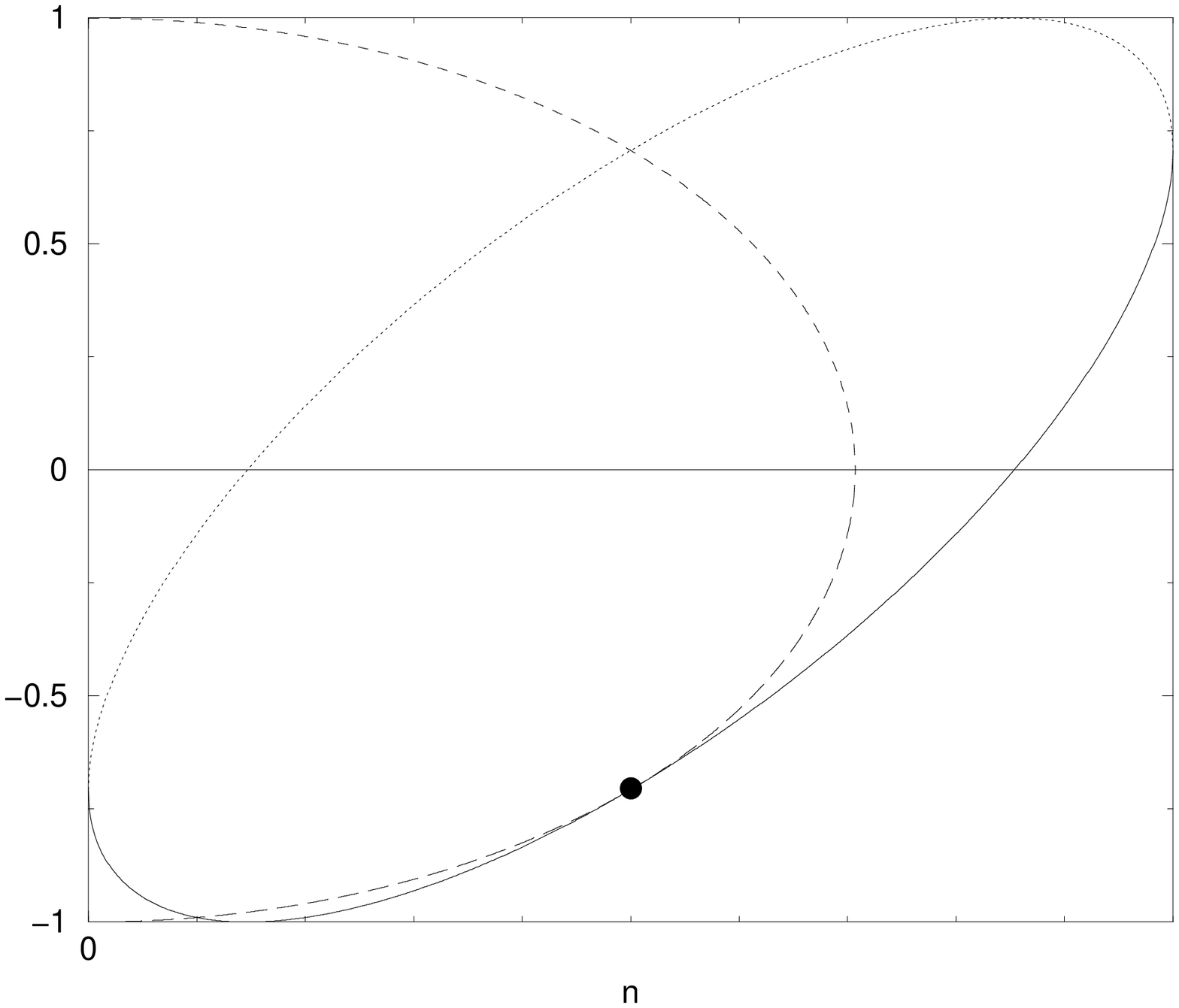, width=5cm, angle=-90 }
\end{center}
\caption{Dependence of different frequencies on $n$ ($0\leq n\leq Q$). 
  Left:  $0<\alpha<\pi/2$, right: $\pi/2<\alpha  <\pi$.
  $\omega_n$ is represented by solid line,
  $\omega_n^*$ by  dotted line, $\widetilde{\omega}_n$ by dashed line, and
  $-\widetilde{\omega}_n$  by long-dashed line. The full small circles 
  indicate the point with coordinates $(Q\sin^2\alpha, \cos \alpha)$ where two
  different frequencies (with $\varphi=0$) coincide.}
\label{omega12}
\end{figure}
In the first region one should expand the factors near 
$Q\sin^2\alpha$, i.e.
\begin{equation}
n'=Q\sin^2 \alpha +\delta n,
\end{equation}
where 
\begin{equation}
\delta n=-\{Q\sin^2 \alpha\}+q
\end{equation}
with integer $q$ and the similar expansion for $m'$. (As above $\{x\}$
denotes the fractional part of $x$.)

The required series  when $0<\alpha <\pi/2$ are easily obtained from
Eqs.~(\ref{omegasmall1})-(\ref{tildesmall})
\begin{equation}
\omega_n^*|_{n=n^*+\delta n}= \cos \alpha-
\frac{\delta n}{Q\cos \alpha }- 
\frac{(\delta n)^2}{4 Q^2 \sin^2 \alpha \cos^3 \alpha } +
{\cal O}\left ( (\frac{\delta n}{Q})^3 \right ),
\label{seriesomega1}
\end{equation}
and
\begin{equation}
\widetilde{\omega}_m|_{m=n^*+\delta m}= |\cos \alpha|-
\frac{\delta m}{Q|\cos \alpha |}- 
\frac{(\delta m)^2}{2 Q^2 \sin^2 \alpha |\cos^3 \alpha |} +
{\cal O}\left ((\frac{\delta m}{Q})^3 \right ).
\label{seriesomega2}
\end{equation} 
For completeness we add 
\begin{equation}
\omega_n|_{n=n^*+\delta n} =\cos \alpha (1-4\sin^2 \alpha )+
{\cal O}(\frac{\delta n}{Q}).
\end{equation}
At small $n\ll Q$
\begin{equation}
\omega_n =\cos \alpha-2\sin \alpha \sqrt{\frac{n}{Q}}+
{\cal O}(\frac{n}{Q}),\;\;
\omega_n^{*} = \cos \alpha+2\sin \alpha \sqrt{\frac{n}{Q}}+
{\cal O}(\frac{n}{Q})
\label{seriessmalln}
\end{equation}
In the first region when $n$ is small and $0<\alpha <\pi/2$ one formally  has
\begin{equation}
\prod_{m'}(\omega_n-\widetilde{\omega}_{m'})\approx \prod_q 
\left (
  -2\sin \alpha \sqrt{\frac{n}{Q}}+\frac{-\{Q\sin^2 \alpha\}+q}{Q\cos \alpha}
\right ).
\label{prod1}
\end{equation}
Similarly in the second region the expansion of $\omega_n-\omega_{n'}^*$ gives
\begin{equation}
\prod_{n'}(\omega_n-\omega_{n'}^*) \approx 
\prod_q \left 
(-2\sin \alpha \sqrt{\frac{n}{Q}}+\frac{-\{Q\sin^2 \alpha\}+q}{Q\cos \alpha}
\right ).
\label{prod2}
\end{equation}
The two products (\ref{prod1}) and (\ref{prod2}) are identical and cancel
each other in the expression for the reflection coefficient. The same
remains true when $\pi/2<\alpha<\pi$.
Therefore a dominant contribution comes only from the region of small $n'$ 
where
\begin{eqnarray}
\prod_{n'\neq n}\frac{\omega_n-\omega_{n'}^*}{\omega_n-\omega_{n'}}&\approx&
\prod_{n'\neq n}^N\frac{1+\sqrt{\frac{n}{n'}}}{1-\sqrt{\frac{n}{n'}}}
\nonumber\\
&\times& \exp (-\int_0^N [\log (1+\sqrt{\frac{n}{n'}}) -
\log (1-\sqrt{\frac{n}{n'}})]dn').
\end{eqnarray}
The last term is added because the total integral is canceled by other terms.
When $N\rightarrow \infty$ the integral equals 
$2\sqrt{n}\sqrt{N} +{\cal O}(1/N)$.

Introducing the convergence factors one gets that this product tends to the
following function
\begin{equation}
f_n=\exp \left (2\sqrt{n}(\sum_{n'\neq n
  }^N\frac{1}{\sqrt{n'}}-\sqrt{N}/2)\right )
\prod_{\begin{array}{c}{\scriptstyle n'\neq n}\\ 
{\scriptstyle n'\neq 0}\end{array} }^N\frac{1+\sqrt{\frac{n}{n'}}}
{1-\sqrt{\frac{n}{n'}}}e^{-2\sqrt{\frac{n}{n'}}}.
\end{equation}
When $N\to \infty$ using  Eq.~(\ref{zeta}) we obtain 
\begin{equation}
f_n=e^{2\sqrt{n}\zeta ({\scriptstyle \frac{1}{2}})-2}
\prod_{\begin{array}{c}{\scriptstyle n'\neq n}\\ 
{\scriptstyle n'\neq 0}\end{array} }^{\infty}\frac{1+\sqrt{\frac{n}{n'}}}
{1-\sqrt{\frac{n}{n'}}}e^{-2\sqrt{\frac{n}{n'}}}.
\label{fn}
\end{equation}
Combining all factors together one concludes that in the limit 
$Q\rightarrow \infty$ the first term of the expansion of the modulus of the 
reflection coefficient at small $n\neq 0$   has the following asymptotics
\begin{equation}
|R_n|^2\stackrel{\varphi \to 0}{\to}  Q\varphi^2 |r_n|^2 =4u^2 |r_n|^2,
\label{88}
\end{equation}
where $u$ is defined in (\ref{u}) and
\begin{equation}
|r_n|^2=\frac{|f_n|}{n}.
\label{thrn}
\end{equation}
In Fig.~\ref{fig4} we present the results of numerical calculation of the
first term of expansion of  reflection coefficient from Eq.~(\ref{modrn}) 
together with asymptotic formula (\ref{thrn}) for this quantity. The
agreement  becomes better at larger $Q$ as it should be.

\begin{figure}
\begin{center}
\epsfig{file=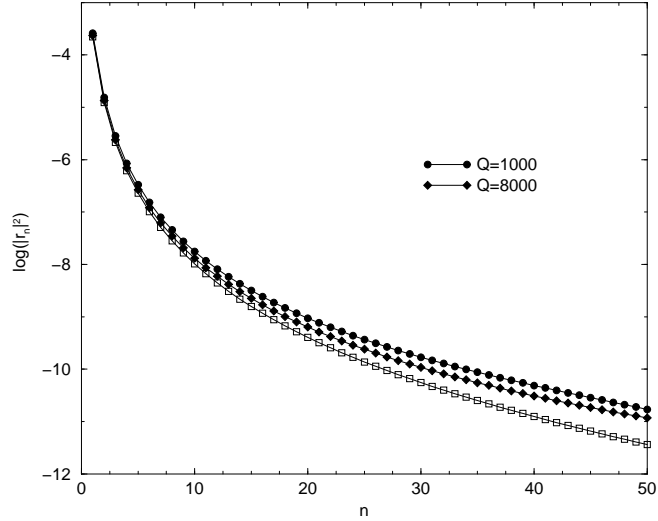, width=7cm, angle=-90 }
\end{center}
\caption{ The first term of the expansion of the reflection coefficient
  at small $n\geq 1$ into a series of $u$. 
  Two upper curves are  $|r_n|^2$
  computed at $Q=1000$ and $Q=8000$ with $\alpha =\pi/3$. The lower curve is
  the asymptotic expression (\ref{thrn}) for this quantity.}
\label{fig4}
\end{figure}

Notice that (when $\alpha$ is not too close to 0, $\pi/2$, and $\pi$) the
reflection coefficient at small $n$ do not depend on $\alpha$. One can check
that when $Q\to \infty$ the transmission coefficients with small $m$ are always negligible.

\section{Reflection and transmission coefficients at \\large $n$ and
  $\varphi\to 0$}\label{largen}

One can prove that for large $n$ the only important contribution
comes from $n$  close to $n^*=Q\sin^2 \alpha$ where two different
frequencies coincide (see Fig.~\ref{omega12}). When $0<\alpha<\pi/2$ 
in this region  only the transmission is noticeable and when 
$\pi/2<\alpha<\pi$ only the reflection is important. 

Let us consider first the case $0<\alpha<\pi/2$. The expression (\ref{modtm})
for  the transmission coefficient modulus $|T_m|^2$ can conveniently be  
rewritten in the following form explicitly separating small factors 
\begin{eqnarray}
 |T_m|^2&=&\sin^2 \varphi \left |\frac{2 (\widetilde{\omega}_m-\omega_m^*)}
{(\widetilde{\omega}_m-\omega_0)(\omega'-\omega_m^*)\widetilde{\omega}_m}
\prod_{n'\neq m} \frac{\widetilde{\omega}_m-\omega_{n'}^*}{\omega'-\omega_{n'}^*}
\prod_{m'\neq m}\frac{\omega'-\widetilde{\omega}_{m'}}
{\widetilde{\omega}_m-\widetilde{\omega}_{m'}}\right .\nonumber\\
&\times& 
\left .\prod_{n'}\frac{\omega'-\omega_{n'}}{\widetilde{\omega}_{m}-\omega_{n'}}
\prod_{m'}\frac{\widetilde{\omega}_m+\widetilde{\omega}_{m'}}{\omega'+\widetilde{\omega}_{m'}}
\right |.
\end{eqnarray}
In these formulas $n'$ and $m'$ are non zero and we assume that
$m=Q\sin^2\alpha +\delta m$ where $\delta m$ is of the order of $\sqrt{Q}$.

From Eqs.~(\ref{seriesomega1}), (\ref{seriesomega2}), and
(\ref{seriessmalln}) formal expansions for the above products can be obtained.
The product over $n'$ contains terms with small $n'=n$ and large
$n'=Q\sin^2\alpha +\delta n'$. Putting $\delta n'=\delta m+q$ and taking 
into account dominant terms one gets  
\begin{eqnarray}
\prod_{n'\neq m} (\widetilde{\omega}_m-\omega_{n'}^*)
&\approx& \prod_{n\geq 1} \left (-\frac{\delta m}{Q\cos \alpha} -
2\sin \alpha \sqrt{\frac{n}{Q}}\right )\nonumber \\
&\times&\prod_{q\neq 0}
\left (\frac{q}{Q\cos \alpha}-\frac{(\delta m)^2}
{4Q^2 \sin^2 \alpha \cos^3 \alpha }\right ).
\end{eqnarray}
The first product comes from small $n$ and the second one is due to the
contributions with large $n'=Q\sin^2\alpha +\delta m+q$.

Computing other products in the similar way one obtains that the first term
of the expansion of the transmission coefficient into a series of $u$ has
the form
\begin{equation}
|T_m|^2\stackrel{u\to 0}{\to}\frac{8u^2}{Q\sin^2 2\alpha}g(u_f),
\label{rdtm}
\end{equation}
where the scaled variable $u_f$  is related to $m$ as follows
\begin{equation}
u_f=\frac{m-Q\sin^2 \alpha}{\sqrt{Q}\sin 2\alpha},
\label{um}
\end{equation}
and the function $g(x)$ is 
\begin{equation}
g(x)=e^{2\zeta({\scriptstyle \frac{1}{2}})x}\prod_{n=1}^{\infty} 
\left (1+\frac{x}{\sqrt{n}}\right )^2
\left (1+\frac{x^2}{n}\right )e^{-\frac{2x}{\sqrt{n}}}.
\label{gx}
\end{equation}
The renormalization factor $e^{2\zeta({\scriptstyle \frac{1}{2}})x}$ has been computed exactly
as it was done above for the reflection coefficient at small $n$.

When $\pi/2<\alpha <\pi$ one can similarly show that transmission coefficients
at small $\varphi$ are  negligible and the reflection coefficients with $n$
close to $n=Q\sin^{2} \alpha$ differ only by the factor $1/2$ from the
transmission ones with the same value of $n$ and $\alpha=\pi -\alpha$
\begin{equation}
|R_n(\alpha)|^2\stackrel{u\to 0}{\to}\frac{1}{2}|T_n(\pi-\alpha)|^2. 
\label{frn}
\end{equation}
For clarity we explicitly show the dependence of $\alpha$.
The factor $1/2$ with respect to Eq.~(\ref{rdtm}) is related with the 
different normalization of real transmitted modes (\ref{transition}) and 
complex reflected ones (\ref{reflectwave}). 

The formulas above are valid when $u_f$ is fixed and $Q\to \infty$. The
meaning of the variable $u_f$ (\ref{um}) is most easily seen when
$\pi/2<\alpha<\pi$. The scattering demi-plans can be considered as a system of
mirrors and the initial ray with $\varphi=0$ after the specular reflection
in these (demi) mirrors will form the angle $2\pi -2\alpha$ with the
scattering plane. For the reflection close to this optical boundary (cf.
(\ref{optbound})) it is naturally to introduce a  new variable $\psi_n$ 
which measures the deviation of the reflected angle from the angle 
$2\pi -2\alpha$ 
\begin{equation}
\varphi_n'=2\pi -2\alpha -\psi_n
\end{equation}
At large $Q$ the variable $u_f$ in (\ref{um}) (with $m=n$) is related with
$\psi_n$ exactly in the same way as the variable $u$ is related to the 
incident angle $\varphi$ (see (\ref{u}))
\begin{equation}
u_f=\sqrt{Q}\sin \frac{\psi_n}{2}.
\label{uf}
\end{equation}
We write $u_f$ without the index $n$ in order to stress that for  large
angle scattering the discreetness of reflected and transmitted waves plays
no role and the asymptotic formulas (\ref{rdtm}), (\ref{frn}) can
be considered as functions of a continuous variable $u_f$. 

In Fig.~(\ref{fig6}) we present the results of numerical calculations of the
function $g(u_f)$ calculated from Eq.~(\ref{modtm}) together with the 
asymptotic formula (\ref{rdtm}).
The agreement is quite good. The reflection coefficient for
$\pi/2<\alpha<\pi$ is also well described by the asymptotic relation 
(\ref{frn}).  
\begin{figure}
\begin{center}
\epsfig{file=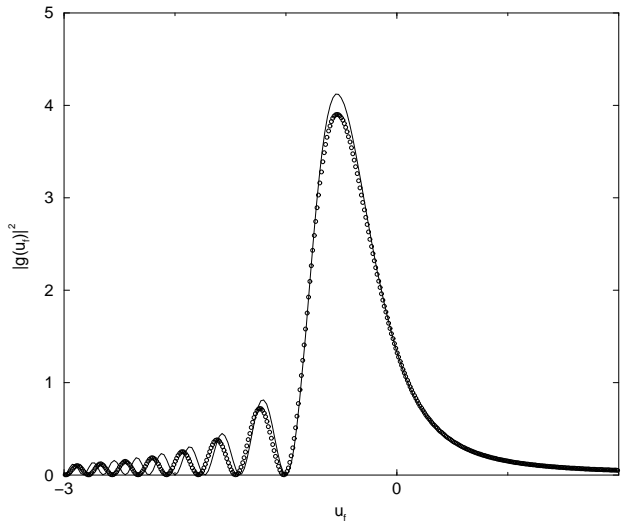, width=7cm, angle=-90 }
\end{center}
\caption{The first term of the expansion of the transmission coefficient
  for $Q=1000$ and $\alpha=\pi/3$ into a series of $u$ (dots).
   The solid line is asymptotic formula (\ref{gx})}
\label{fig6}
\end{figure}

\section{Reflection and transmission coefficients at finite $u$}\label{finiteu}

Eqs.~(\ref{88}), (\ref{rdtm}) and (\ref{frn})  represent the first terms of 
the expansion of the transmission and reflection coefficients in the power of
$u=\sqrt{Q}\sin (\varphi/2)$. 

The purpose of this Section is to calculate these coefficients at finite
values of $u$. The calculations are performed exactly as in the precedent 
Sections. First we note that the frequencies $\omega_n$ and $\omega_n^*$
depend on $u$ only in the combination  $n(u)=n+u^2$. Therefore precedent
formulas remain valid 
when $n$ is substituted by $n(u)=n+u^2$ and the dominant contributions
come from the same terms as above. We omit the details and present only 
the final answers.

The reflection coefficients at all small $n$ (for negative $n\geq -[u^2]$,
$n=0$, and positive $n$) are given by 
\begin{equation}
|R_n(u)|^2=\left (\frac{2u}{u+\sqrt{n+u^2}}\right )^2G(\sqrt{n+u^2})G(u)
\label{rnfinal}
\end{equation}
and the function $G(x)$ has the following form
\begin{equation}
G(x)=e^{2x\zeta ({\scriptstyle \frac{1}{2}})}
\prod_{n'\geq 1}e^{-2\frac{x}{\sqrt{n'}}}\prod_{ \begin{array}{c} 
  {\scriptstyle n'\geq 0}\\{\scriptstyle n'\neq [x^2]}\end{array}}
\left |\frac{1+\frac{x}{\sqrt{n'+\{u^2\}}}}{1-\frac{x}{\sqrt{n'+\{u^2\}}}}
\right |
\label{G}
\end{equation}
The terms with the same value of $n'$ should be grouped together and then 
the total product converges.

In Fig.~\ref{fig7} the results of numerical calculations for
$|R_n(u)|^2$ with $n=-3,\ldots, 3$,  $Q=8000$ and $\alpha=\pi/3$ are compared 
with the asymptotic formula (\ref{rnfinal}) and the excellent agreement is 
found.
\begin{figure}
\begin{center}
\epsfig{file=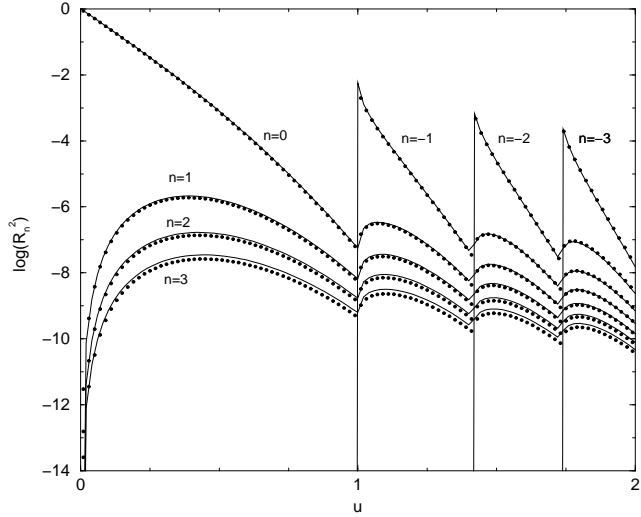, width=7cm, angle=-90 }
\end{center}
\caption{ The reflection coefficients with $n=-3,\ldots ,3$ with  $Q=8000$
  and  $\alpha=\pi/3$ (solid lines) together with  the asymptotic formula 
  (\ref{rnfinal}) (dots) for the same values of $n$.}
\label{fig7}
\end{figure}

In the discussion of the scattering at large angles corresponding to $n$ 
close to $Q\sin^2\alpha $
it is convenient to introduce reduced transmission  $t(u_f,u)$ and
reflection $r(u_f,u)$ coefficients in the following way  
\begin{equation}
T_m=\frac{2}{\sqrt{Q}\sin 2\alpha}t(u_f,u),
\label{redtm}
\end{equation}
and 
\begin{equation}
R_n=\frac{2}{\sqrt{Q}\sin 2\alpha}r(u_f,u),
\label{redrn}
\end{equation}
where the variable $u$ defined by Eq.~(\ref{u}) fixes the initial angle and the 
variable $u_f$ is related with the scattering angle by Eq.~(\ref{um}).

As above one can demonstrate that when $0<\alpha<\pi/2$ only the
transmission coefficients are important and when $\pi/2<\alpha <\pi$ the
reflection coefficients dominate and asymptotically in corresponding regions
\begin{equation}
|r(u_f,u)|^2=\frac{1}{2}|t(u_f,u,)|^2=g(u_f,u)
\label{finalrn}
\end{equation}
where
\begin{eqnarray}
g(u_f,u)&=&u^2 e^{2(u+u_f)\zeta({\scriptstyle \frac{1}{2}})}
\prod_{n'\geq 1}e^{-2\frac{u+u_f}{\sqrt{n'}}}
\prod_{\begin{array}{c}
  {\scriptstyle  n'\geq 0}\\{\scriptstyle n'\neq [u^2]}\end{array}}
\left |\frac{1+\frac{u_f}{\sqrt{n'+\{u^2\}}}}{1-\frac{u}{\sqrt{n'+\{u^2\}}}}
\right |^2\nonumber\\
&\times& \prod_{n'\geq [u^2]+1} \left (1+\frac{u_f^2-u^2}{n'} \right ).
\label{tmfinal}
\end{eqnarray}
As above, the terms with the same $n'$ should be combined together for 
convergence.

The reduced coefficients $t(u_f,u)$ and $r(u_f,u)$ introduced in
Eqs.~(\ref{redtm}), (\ref{redrn}) can be considered as 
the reflection and transmission coefficients for the scattering in the
(continuous) interval of `angles' $u_f$ as it follows from the current 
conservation (\ref{current}) rewritten in terms of  new variables 
\begin{equation}
u=\sum_{n\geq -[u^2]}\sqrt{n+u^2}|R_n(u)|^2+
2\int_{-\infty}^{\infty}du_fg(u_f,u).
\end{equation}
Here $R_n(u)$ mean the reflection coefficients at small $n$. 

When $u \to \infty$ the precedent equations can considerably be simplified
by noting that  in this limit the products in these equations are large when
$u_f$ is close to $-u$. Putting $u_f=-u+\delta$ and taking into account the
dominant terms one gets  
\begin{equation}
g(u_f,u)\approx \left (\frac{\sin (2\pi (u_f+u)u)}{2\pi (u+u_f)}\right )^2.
\label{approx}
\end{equation}
Practically this approximation works well even with $u\geq 1$.
The semiclassical derivation of this expression is performed in
Section~\ref{kirchhoff}. 

In Fig.~\ref{fig8} we present numerically computed transmission coefficients with $u=2.1$ and
$\alpha=\pi/3$ and reflection coefficients with $\alpha=2\pi/3$ 
for $Q=1000$, $Q=4000$, and $Q=16000$ together with the 
asymptotic formulas (\ref{tmfinal})  and (\ref{finalrn}). With increasing 
$Q$ the agreement becomes better and better. The approximation
(\ref{approx}) is hardly distinguished from the exact asymptotic formula
(\ref{tmfinal}).  

\begin{figure}
\begin{center}
\epsfig{file=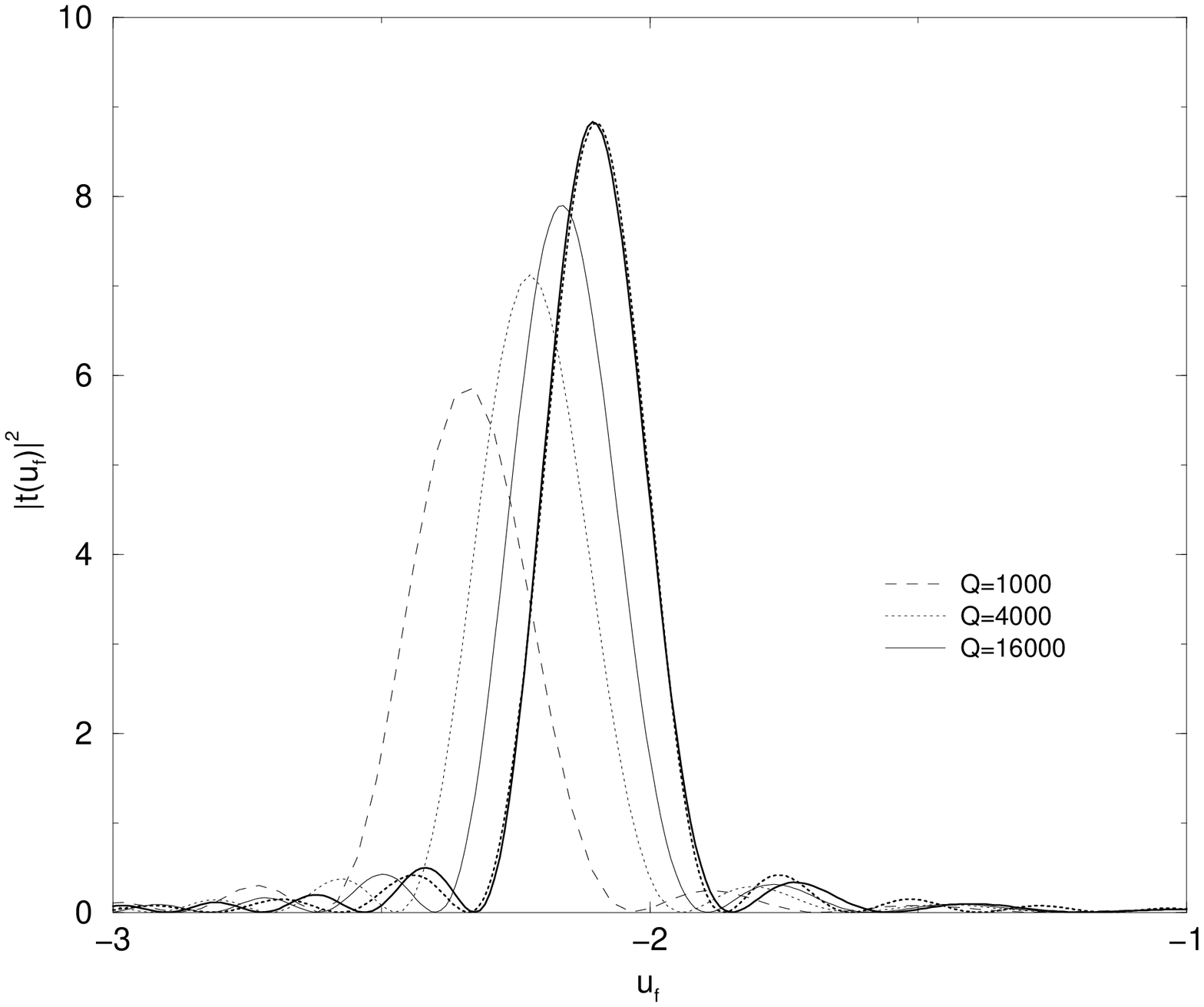, width=5cm, angle=-90 }
\epsfig{file=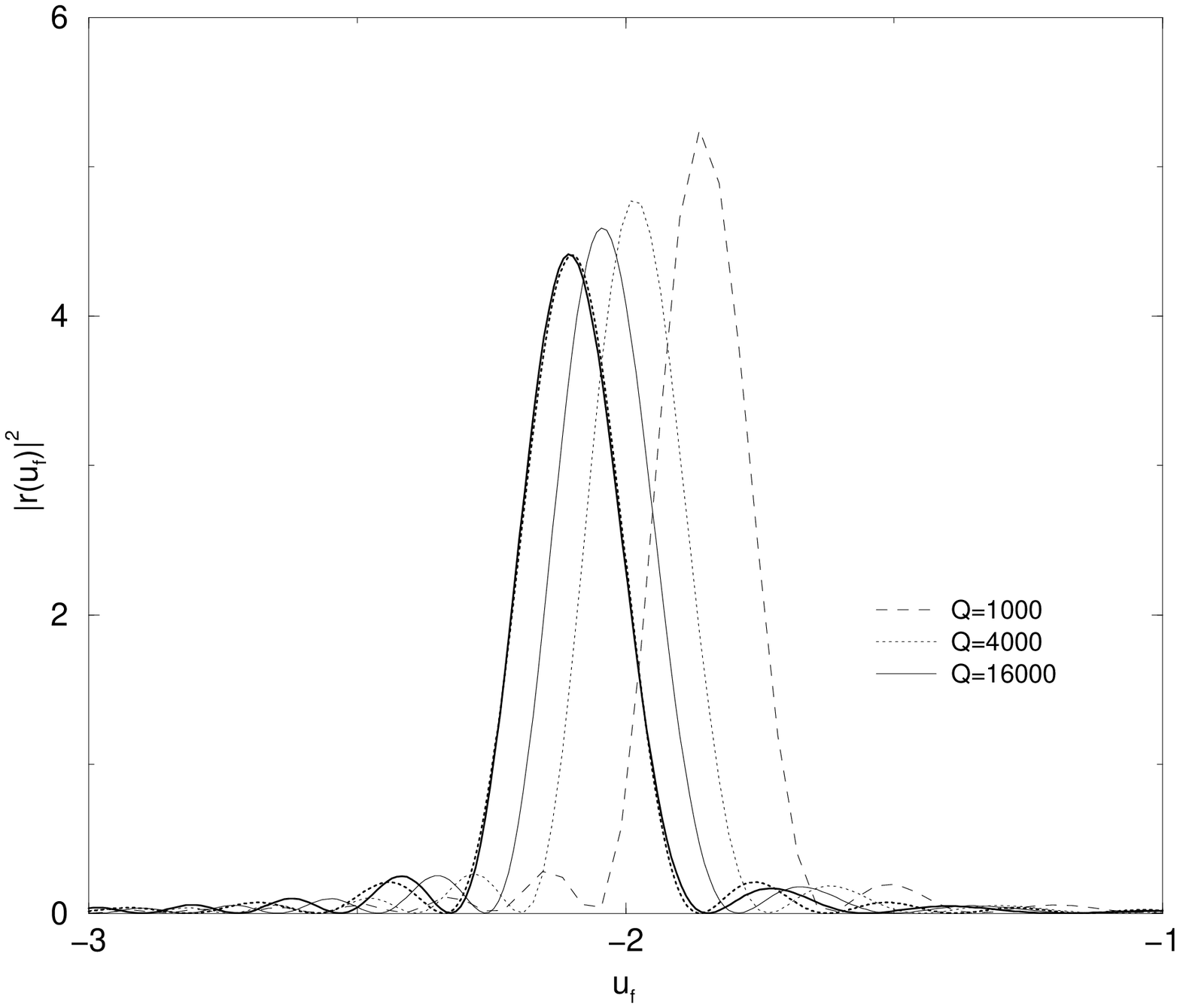, width=5cm, angle=-90 }
\end{center}
\caption{ Reduced transmission coefficients with   $\alpha=\pi/3$
(left) and the reduced reflection coefficients  with   $\alpha=2\pi/3$ (right).
In both figures $u=2.1$. Dashed line: $Q=1000$, dotted line:  $Q=4000$,
  solid line:  $Q=16000$.  
  The thick solid line are  the asymptotic formulas (\ref{tmfinal})  and
  (\ref{finalrn}).
  The thick dotted line is the approximation (\ref{approx}).}
\label{fig8}
\end{figure}
 
\section{Scattering when $\alpha=\pi/2$}\label{pi2}

The formulas of the precedent Sections are not valid when $\alpha=\pi/2$,
i.e. when the demi-plans are perpendicular to the scattering plane.
Nevertheless, asymptotic expressions  for large $Q$ can also be obtained in
this case but the results depend on the fractional part of $Q$ and, strictly
speaking, the semiclassical limit $Q\to \infty$ does not exist. The main
reason for such behaviour is related with the existence of a trivial solution
for some particular values of $Q$. It is easily seen that when $\alpha=\pi/2$ 
the following function (in the notation of Fig.~\ref{fig3}) 
\begin{equation}
\Psi^{(0)}(x,y) =\sin (kx\cos \varphi ) e^{ik z\sin \varphi }
\label{exact}
\end{equation}
is an exact solution of our problem (i.e. it vanishes at the demi-plans) 
provided 
\begin{equation}
kd\cos \varphi =\pi l,
\end{equation}
with integer $l$. In the notations (\ref{Q}) and (\ref{u}) this
relation takes the form
\begin{equation}
\{Q-2u^2\}=0,
\label{fractpart}
\end{equation}
where $0\leq \{x\}<1$ is the fractional part of $x$.

Irrespectively how large is the dimensionless momentum $Q$, close to 
points when Eq.~(\ref{fractpart}) is fulfilled, the appearance and 
disappearance of the exact solution (\ref{exact}) strongly perturb the 
reflection and transmission coefficients.

Though the limit $Q\to \infty$ does not exist, a special limiting case
when $\{Q\}$ fixed and $[Q]\to \infty$ can be computed from the approach 
developed in the preceding Sections. We omit the details and present only the 
final formulas.

Let us define a function $W(t)$ as follows.
\begin{eqnarray}
&&W(t)=e^{2(2-\sqrt{2})t\zeta({\scriptstyle \frac{1}{2}})}
\prod_{n'\geq 1}e^{-2(2-\sqrt{2})\frac{t}{\sqrt{n'}}}
\nonumber\\
&&\times \prod_{n'\geq 0}\ ^\prime
\left (\frac{1+\frac{t}{\sqrt{n'+\{u^2\}}}}{1-\frac{t}{\sqrt{n'+\{u^2}\}}}\right )
\left (\frac{1+\frac{t}{\sqrt{n'+\{Q-u^2\}}}}{1-\frac{t}{\sqrt{n'+\{Q-u^2\}}}}\right )\\
&&\times
\left (\frac{1-\frac{\sqrt{2}t}{\sqrt{n'+\{Q\}}}}{1+\frac{\sqrt{2}t}
    {\sqrt{n'+\{Q\}}}}\right ).
\nonumber  
\label{Wx}
\end{eqnarray}
We shall need this function at special values of the arguments
$t=\sqrt{n+\{u^2\}}$, $t=\sqrt{n+\{Q-u^2\}}$ and
$t=-\frac{1}{\sqrt{2}}\sqrt{n+\{Q\}}$ with integer $n$.
The prime in the second product in Eq.~(\ref{Wx}) means that when
$t=\sqrt{n+\{u^2\}}$ the term with $n'=n$ is omitted in the first factor,
when  $t=\sqrt{n+\{Q-u^2\}}$ the term with $n'=n$ is omitted in the second
factor and when $t=-\frac{1}{\sqrt{2}}\sqrt{n+\{Q\}}$ the term with $n'=n$
is absent in  the third factor.  

The reflection coefficients at small $n$ 
\begin{equation}
|R_n(u)|^2=\left (\frac{2 u}{u+u_n}\right )^2W (u_n)W(u),
\label{108}
\end{equation}
where $u_n=\sqrt{n+u^2}$ determines the allowed value of small reflection 
angle (see (\ref{uprime}) and (\ref{nu})).

The reflection coefficients at large  $n=[Q]-q$ is given by the same
expression (\ref{108}) but $u_n$ is substituted by $u_q$  
\begin{equation}
|R_q(u)|^2=\left (\frac{2 u}{u+u_q}\right )^2
W (u_q)W(u),
\label{109}
\end{equation}
where
\begin{equation}
u_q=\sqrt{q+\{Q\}-u^2}
\end{equation}
has the meaning of the allowed value of small deviation of reflected angle
from its maximum possible value. It means that if one writes
$\varphi_n'=\pi-\delta \varphi_n'$ then for $n=[Q]-q$ with large $[Q]$ and
fixed $q$ 
\begin{equation}
u_q\approx \sqrt{Q}\sin \frac{\varphi_{[Q]-q}'}{2}.
\end{equation}
The reflection coefficient $R_q$ with integer $q=0,1,\ldots ,$ exists for 
a finite interval of $u$ such that $0\leq u\leq \sqrt{q+\{Q\}}$.  

The transmission coefficients for $m=[Q]-p$ with $p=0,1,\ldots ,$ are given
by the same expression (\ref{108}) but with the substitution $u_n\to u_p$
\begin{equation}
|T_p(u)|^2=\left (\frac{2 u}{u+u_p}\right )^2
W (u_p)W(u),
\label{110}
\end{equation}
where
\begin{equation}
u_p=-\frac{1}{\sqrt{2}}\sqrt{p+\{Q\}}.
\end{equation}
If for $\alpha =\pi/2$ one determines the angle of transmission $\phi_m$ from the natural
relation   $\cos \phi_m=m/Q$ then for  $m=[Q]-p$ $u_p$ is related with
$\delta \phi_p=\pi-\phi_{m=[Q]-p}$ in the same way as above
\begin{equation}
u_p\approx \sqrt{Q}\sin \frac{\delta \phi_{[Q]-p}}{2}.
\end{equation}

Ay Figs.~\ref{fig11}-\ref{fig13} we present the the results of numerical 
calculations of the reflection and transmission coefficients together 
with the above asymptotic formulas. The agreement is very good.

\begin{figure}
\begin{center}
\epsfig{file=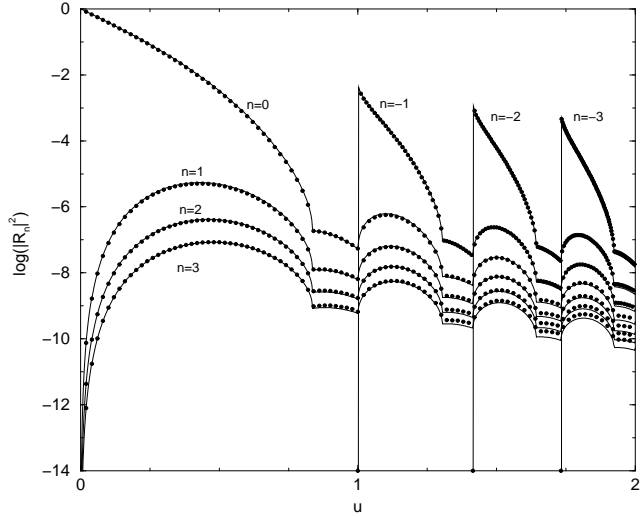, width=7cm, angle=-90 }
\end{center}
\caption{Reflection coefficients with $n=-3,\ldots ,3$ with
  $\alpha=\pi/2$ and $Q=8000.7$ (solid lines) together with the asymptotic
  formula (\ref{108}) (dots) for the same values of $n$ and $\{Q\}$.}
\label{fig11}
\end{figure}

\begin{figure}
\begin{center}
\epsfig{file=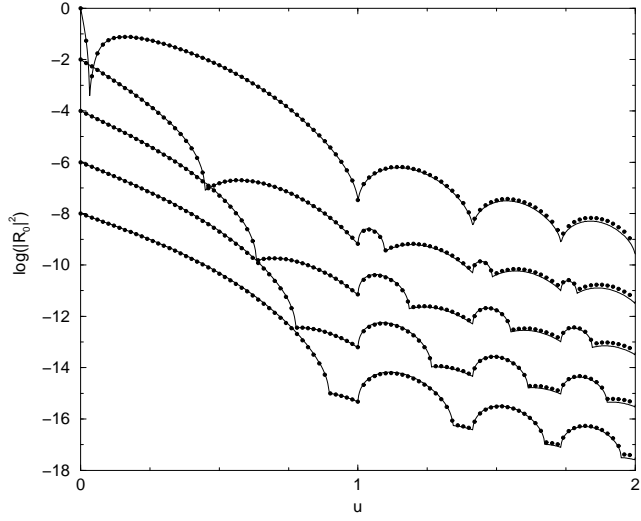, width=7cm, angle=-90 }
\end{center}
\caption{ Elastic reflection coefficient with $\alpha=\pi/2$ and
  $[Q]=8000$ for different values of fractional part of $Q$ (solid lines). 
  From top to bottom 
  $\{Q\}=.001, 0.201, 0,401, 0.601, 0.801$. For clarity each curve is
  lowered with respect to the precedent by 2 units. Dots represent the
  asymptotic formula (\ref{108}).}
\label{fig12}
\end{figure}

\begin{figure}
\begin{center}
\epsfig{file=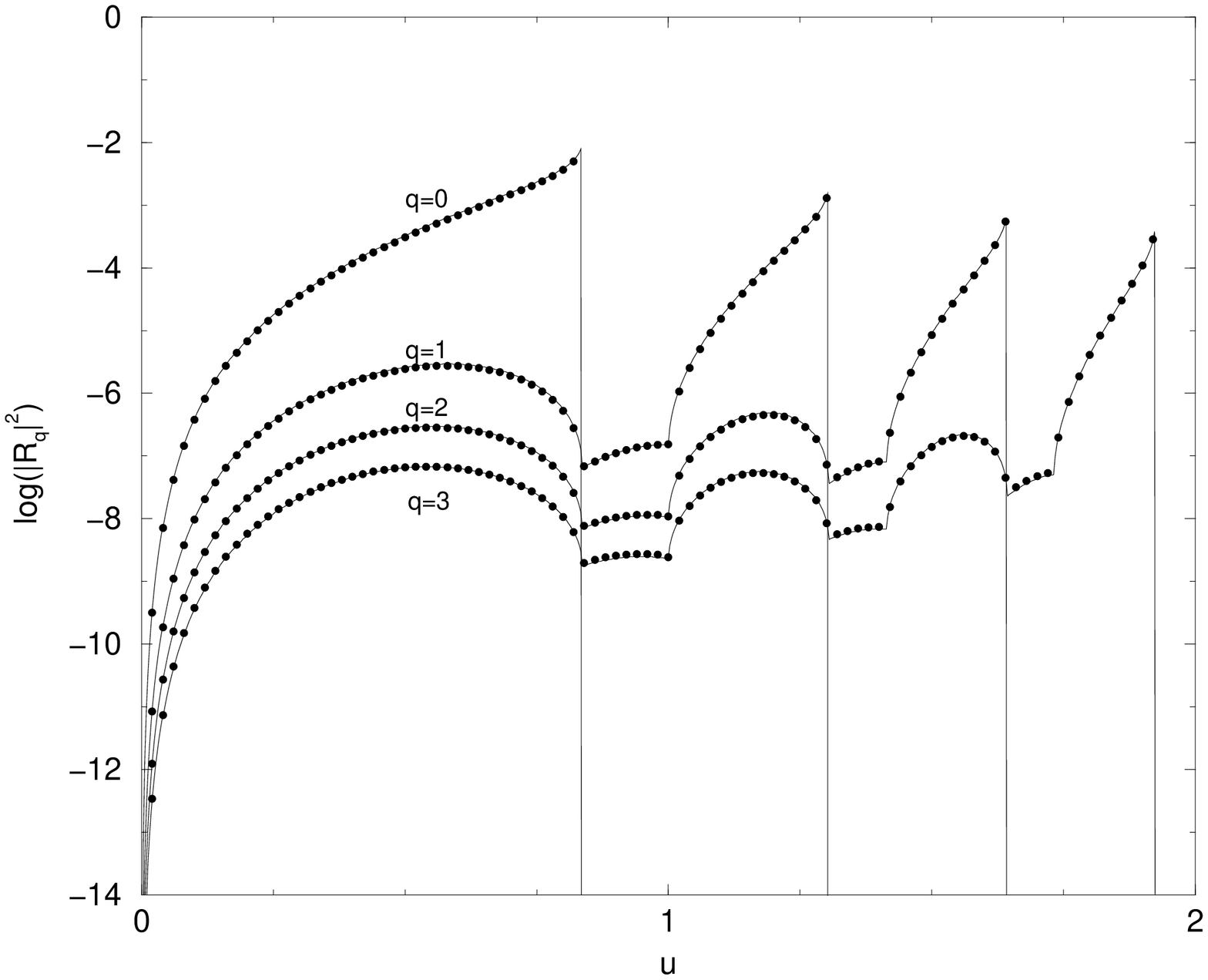, width=7cm, angle=-90 }
\end{center}
\caption{Reflection coefficients with $\alpha=\pi/2$ and
  $Q=8000.7$ for the largest values of $n\equiv [Q]-q$ with $q=0,\ldots ,3$ 
  (solid lines). Dots represent the  asymptotic formula (\ref{109}).}
\label{fig13}
\end{figure}

\begin{figure}
\begin{center}
\epsfig{file=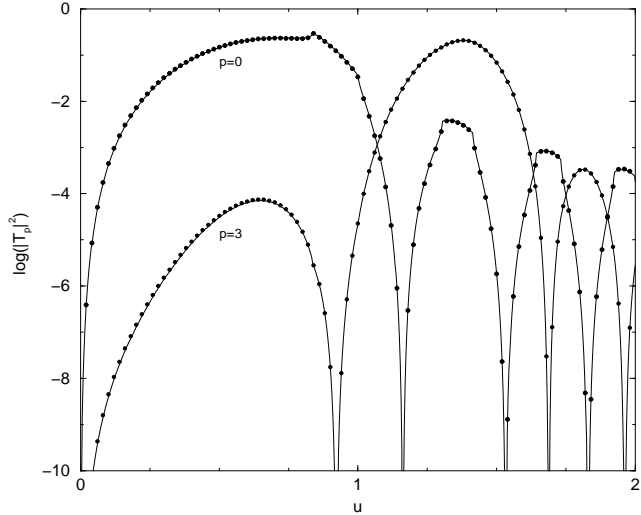, width=7cm, angle=-90 }
\end{center}
\caption{ Transmission coefficients with $\alpha=\pi/2$ and
  $Q=8000.7$ for the largest values of $m\equiv \{Q\}-p$ with $p=0$ and $p=3$ 
  (solid lines). Dots represent the  asymptotic formula (\ref{110}).}
\label{fig14}
\end{figure}

\section{Kirchhoff approximation}\label{kirchhoff}

Semiclassical limit of different cases of multiple diffraction near the
optical boundaries have been considered in Ref.~\cite{BPS}. 
In particular in this paper 
the contribution to the trace formula from diffractive orbits
close to $n$-fold repetition of a primitive diffractive orbit has been
calculated.

In the Kirchhoff approximation this contribution is given by the diagram of
Fig.~\ref{fig1}.
\begin{figure}
\begin{center}
\epsfig{file=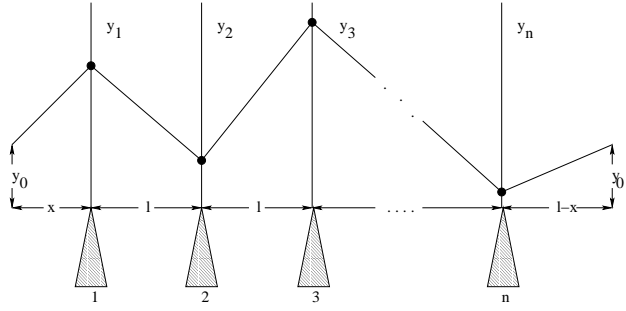, width=7cm, angle=-90 }
\end{center}
\caption{ Schematic representation of multiple diffraction near optical
  boundary of $n$ equally spaced obstacles.}
\label{fig1}
\end{figure}
Each line in this figure corresponds to the free Green function
\begin{equation}
G_0(\vec{x},\vec{x}')=\frac{e^{ikl-3\pi i/4}}{\sqrt{8\pi k l}},
\end{equation}
where $l=|\vec{x}-\vec{x}'|$ is the distance between two points $\vec{x}$,
$\vec{x}'$ and $k=\sqrt{E}$ is the momentum. Each circle at Fig.~\ref{fig1} 
describes the
convolution of two Green functions and in the Kirchhoff approximation it
gives the factor $-2ik$. The role of the obstacles (corners) consists in the
restriction of the integration over $y_i$ to the half line. 

In Ref.~\cite{BPS} it was demonstrated that the contribution to the trace
formula from such trajectories has the form
\begin{equation}
\rho^{(diff)}(E)=-\frac{l}{16\pi k} A_n e^{ikln}+c.c. ,
\label{diff}
\end{equation}
where $A_n$ is given by the following $n$-fold integral
\begin{equation}
A_n=\frac{4e^{-i\pi n/4}}{\pi^{n/2}}
\int_0^{\infty}ds_1 \left [
\int_{-\infty}^{\infty}ds_2\ldots \int_{-\infty}^{\infty}ds_n
-\int_{0}^{\infty}ds_2\ldots \int_{0}^{\infty}ds_n\right ]e^{i\Phi(\vec{s}\ )}
\end{equation}
with
\begin{equation}
\Phi(\vec{s}\ )=(s_1-s_2)^2+(s_2-s_3)^2+\ldots +(s_{n-1}-s_{n})^2+(s_n-s_1)^2.
\end{equation}
The presence  a discrete symmetry in  this quadratic form permits the
analytical computation of this integral and \cite{BPS}
\begin{equation}
A_n=\frac{1}{\pi}\sum_{q=1}^{n-1}\frac{1}{\sqrt{q(n-q)}}.
\label{an}
\end{equation}
When $n\rightarrow \infty$ the sum over $q$ can be substituted by the
integral and 
\begin{equation}
\lim_{n\rightarrow
  \infty}A_n=\frac{1}{\pi}\int_0^{n}\frac{dq}{\sqrt{q(n-q)}}=1
\label{limit}
\end{equation}
and as it was noted in Ref.~\cite{BPS} the contribution (\ref{diff}) in this
limit coincides with the contribution from the boundary trajectory 
reflected from a straight mirror with the Dirichlet boundary
condition.

Let us consider this point in details. Assume that after each
reflection with a mirror a trajectory gets a reflection coefficient $R$. 
Then its contribution to the trace formula  computed in the usual manner is
\begin{eqnarray}
\rho^{(boundary)}(E)&=&-2R\frac{e^{ikL-3\pi i/4}}{16\pi\sqrt{k}}  
\int_0^L\frac{dl}{\sqrt{l(L-l)}}\int_0^{\infty}dye^{i2ky^2(1/l+1/(L-l))}
\nonumber\\
&+&c.c.=R\frac{Le^{ikL+i\Phi}}{16 \pi^2 k}+c.c.
\end{eqnarray}
Here $L$ is the length of the trajectory and the factor 2 takes into account
that each  trajectory can be passed in two directions. For simplicity we 
do not consider the symmetry factor and consider the trajectory as being
primitive.

Comparing this result with Eqs.~(\ref{diff}) and (\ref{limit}) one concludes
that these equations can be interpreted as the reflection from a mirror 
with the effective reflection coefficient $R=-1$ as from a boundary  with 
the Dirichlet boundary conditions.

To compute  the behaviour of the reflection coefficient
for large but finite $n$ it is convenient to use Eq.~(\ref{zeta}) from which 
it follows that when $n\rightarrow \infty$
\begin{equation}
A_n \rightarrow 1+\frac{2\zeta({\scriptstyle \frac{1}{2}})}{\pi \sqrt{n}}.
\label{An}
\end{equation}  
One can incorporate this result into the above picture of the reflection
from a straight mirror by assuming that the reflection coefficient $R$
depends on the reflection angle $\varphi$  (or transverse momenta 
$p_y\approx k\varphi$)
\begin{equation}
R(\theta) =-(1+\beta \varphi).
\label{reflection}
\end{equation}
As in the saddle point approximation $\varphi\approx 2y/L$, the only modification 
of the above calculation is the following integral
\begin{equation}
\int_0^{\infty} dy(1+\beta \frac{2y}{L} )e^{2iy^2/L}=
  \frac{\sqrt{\pi L}e^{i\pi/4}}{\sqrt{8k}}
    (1+\beta \frac{\sqrt{2}e^{i\pi/4}}{\sqrt{k\pi L}}).
\end{equation}
Because $L=ln$  one concludes from (\ref{An}) that 
\begin{equation}
\beta=\sqrt{2\frac{kl}{\pi}}e^{-i\pi/4}\zeta(\frac{1}{2}),
\label{beta}
\end{equation}
which up to notations agrees with Eq.~(\ref{2uR0}) obtained by the direct expansion of
the exact solution.

The above considerations demonstrate that small-angle singular reflection 
from a periodic set of demi-plans can be interpreted as much simpler process 
of the specular reflection from a straight mirror whose reflection coefficient at small
angles has behavior as in Eqs.~(\ref{reflection}) and (\ref{beta}). Of
course, at small distances reflection fields for two processes are very 
different but at large distances  they are equivalent.

Similar (but simpler) considerations permit also to understand the
approximation (\ref{approx}) for the reflection (and transmission) at large 
angles.

At Fig.~\ref{fig10} we represent schematically the configuration important
for the large-angle scattering. The rays reflect from small  parts of the
scattering demi-plans restricted by the indicated corners. The width of each
effective mirror is
\begin{equation}
\Delta=d\sin \varphi,
\end{equation}
where, as above, $d$ is the distance between singular corners along the scattering plane
and $\varphi$ is the scattering angle. 
\begin{figure}
\begin{center}
\epsfig{file=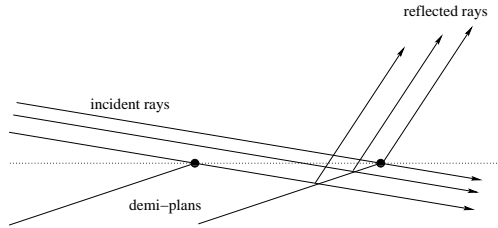, width=7cm, angle=-90 }
\end{center}
\caption{Schematic representation of the scattering at large angles. Black
  circles are the corners of the scattering demi-plans. The dotted line
  represents the scattering plane.}
\label{fig10}
\end{figure}
After unfolding the amplitude of such reflection can be calculated in the Kirchhoff
approximation as the transmission through a slit of the same width by the
usual formula (an additional minus sign is due to the reflection from a
mirror)
\begin{equation}
D(\epsilon)=2ik\int_{-\Delta/2}^{\Delta/2}e^{-iky\sin \epsilon }dy
=\frac{4i}{\sin \epsilon}
\sin \left (\frac{1}{2}k \Delta \sin \epsilon \right ),
\label{DK}
\end{equation}
where $\epsilon$ is the deviation of the scattering angle from the
direction of the specular reflection.

The total reflected field is the sum over all diffracted fields
\begin{equation}
\Psi^{(ref)}(x,y)=\sum_{m=-\infty}^{\infty} e^{ikdm\cos \varphi}D(\delta \varphi')
\frac{e^{ikR_m-3\pi i/4}}{\sqrt{8\pi kR_m}},
\end{equation}
where $R_m=\sqrt{(x-md)^2+y^2}$ is the distance between the $m^{\mbox{th}}$ 
diffractive center and the point of observation with coordinates $(x,y)$.
Using the Poisson summation formula one gets
\begin{equation}
\Psi^{(ref)}(x,y)=\sum_{n=-\infty}^{\infty}\int_{-\infty}^{\infty}dm 
e^{iS(m,n)}D(\epsilon)\frac{e^{-3\pi i/4}}{\sqrt{8\pi kR_m}},
\end{equation}
where
\begin{equation}
S(m,n)=-2\pi mn +k(dm\cos \varphi+R_m).
\end{equation}
When $k\to \infty$ one can use the saddle point method. The saddle point
is obtained from the condition $dS(m,n)/dm=0$ which gives the grating
equation (\ref{grating}). The computation of the second derivative and the
resulting integral leads to
\begin{equation}
\Psi^{(ref)}(x,y)=\sum_{n} \frac{D(\epsilon_n)}{2ikd\sin \varphi_n'}
e^{ik(x\cos \varphi_n'+y\sin\varphi_n')},
\end{equation}
where the reflected angle $\varphi_n'$ is defined by Eq.~(\ref{grating}).

Using Eq.~(\ref{DK}) for the diffraction coefficient of the reflection on a
small mirror and taking into account that in the important region
$\varphi_n'\approx 2\pi -2\alpha$ one obtains the following expression for reflection
coefficient $R_n$
\begin{equation}
R_n=-\frac{2}{kd \sin 2\alpha \sin \epsilon_n}
\sin (\frac{1}{2}kd\sin \varphi \sin \epsilon_n).
\end{equation}
As $\varphi\approx 2u/\sqrt{Q}$ and $\epsilon_n=-2 (u_f+u)/\sqrt{Q}$ with
$Q=kd/\pi$, the last expression equals
\begin{equation}
R_n=-\frac{2}{\sqrt{Q}\sin 2\alpha}\left [\frac{\sin 2\pi u (u_f+u)}{2\pi
  (u_f+u)}\right ],
\end{equation}
whose modulus coincides with Eq.~(\ref{approx}).

\section{Summary}\label{summary}

The exact transmission and reflection coefficients for 
the scattering on infinite number of parallel demi-plans obtained in 
Ref.~\cite{CH} are analyzed in semiclassical limit of large momentum. 

The most interesting (and difficult) case of small incident angle is
considered. More precisely, the incident angle $\varphi$ is chosen in such a
manner that approximately $\varphi=2u/\sqrt{Q}$
where $Q$ is the dimensionless momentum. The limit considered
corresponds to $Q\to \infty$ with fixed $u$.  

It is demonstrated that at small final angles (of the order of $u_n/\sqrt{Q}$)
the transmission is always negligible. The modulus of reflection coefficients
in this case  are independent of $Q$ and the angle $\alpha$ of inclination of
the demi-plans and are given by Eq.~(\ref{rnfinal}).  
The reflection coefficients decay quickly with $n$ (i.e. with increasing of
reflection angle) and $\sum_n \sqrt{n}|R_n|^2$ converges.
The largest  reflection coefficient corresponds to the smallest possible angle
of reflection i.e. $n=-[u^2]$. For very small incident angle the elastic
scattering corresponded to the specular reflection dominates.

The large angle scattering is  noticeable only close to the specular reflection
from the demi-plans.  When $0<\alpha<\pi/2$ only the transmission
is large and the large-angle reflection can be neglected. The transmission
coefficients have  the form
$t(u_f,u)/(\sqrt{Q}\sin 2\alpha)$ where $|f(u,u_f)|^2=2g(u_f,u)$ and
$g(u_f,u)$ is given by Eq.~(\ref{tmfinal}).
Here $u_f$ is the deviation from the angle of mirror reflection 
magnified by the factor $\sqrt{Q}/2$ exactly as $u$ is related with the 
incident angle. For  $\pi/2<\alpha<\pi$ the transmission is small and the
reflection coefficients have similar asymptotics.

The exceptional case of demi-plans perpendicular to the scattering plane
is characterized by the dependence of the fractional part of the momentum.
In the limit $\{Q\}$ fixed and $[Q]\to \infty$ the reflection and
transmission coefficients are independent on $[Q]$ and are given by 
Eqs. ~(\ref{108}) -- (\ref{110}). 

The first two terms of expansion of the exact elastic reflection coefficient
into powers of small incident angle can also be obtained from the results of
Ref.~\cite{BPS} where the Kirchhoff approximation  for multiple
scattering was developed. The large angle scattering is well described by the usual
Kirchhoff approximation when the initial parameter $u\geq 1$.

\end{document}